\documentclass{skaox2006}
\usepackage{graphicx}
\begin{document}
   \title{LOFAR - Opening up a new window on the Universe}

\author{H.J.A. R\"ottgering\inst{1}
\and R. Braun\inst{2} 
\and P. D. Barthel\inst{3}          
\and M. P. van Haarlem\inst{2}
\and G. K. Miley\inst{1}
\and R. Morganti\inst{2,3}
\and \hbox{I. Snellen \inst{1}} 
\and H. Falcke \inst{2,5} 
\and A.G. de Bruyn \inst{2,3}
\and R. B. Stappers \inst{2,4}
\and W.H.W.M. Boland \inst{1} 
\and H.R.  Butcher \inst{2}
\and \hbox{E.J. de Geus \inst{2}}
\and L. Koopmans \inst{3} 
\and R. Fender \inst{4,6} 
\and J. Kuijpers \inst{5} 
\and R.T. Schilizzi \inst{1,7} 
\and C. Vogt \inst{2} 
\and R.A.M.J. Wijers \inst{4} 
\and \hbox{M. Wise \inst{2,4}}  
\and W.N. Brouw \inst{3} 
\and J.P. Hamaker \inst{2} 
\and J.E. Noordam \inst{2} 
\and T. Oosterloo \inst{2} 
\and L. B\"ahren \inst{2,5} 
\and M.A. Brentjens \inst{2,3} 
\and S.J. Wijnholds \inst{2} 
\and J.D.   Bregman \inst{2} 
\and W.A. van Cappellen \inst{2} 
\and A.W. Gunst \inst{2} 
\and G.W. Kant \inst{2} 
\and J. Reitsma   \inst{2} 
\and \hbox{K. van der Schaaf \inst{2} } 
\and C.M. de Vos \inst{2}
}

\institute{Leiden Observatory, University of Leiden,  P.O. 9513, 2300 RA Leiden, The Netherlands
\and
ASTRON,  Dwingeloo, The Netherlands
\and
Kapteyn Instituut,  Groningen, The Netherlands
\and 
Institute Anton Pannekoek, Amsterdam, The Netherlands 
\and 
Department of Astrophysics,  Radboud University, Nijmegen
\and 
School of Physics and Astronomy, University of Southampton, UK 
\and 
SKA Project Office, Dwingeloo, The Netherlands}

 \abstract{

LOFAR, the Low Frequency Array, is a next-generation radio telescope
that is being built in Northern Europe and expected to be fully
operational at the end of this decade.  It will operate at frequencies
from 15 to 240 MHz (corresponding to wavelengths of 20 to 1.2 m). Its
superb sensitivity, high angular resolution, large field of view and
flexible spectroscopic capabilities will represent a dramatic
improvement over previous facilities at these wavelengths. As such,
LOFAR will carry out a broad range of fundamental astrophysical
studies.

The design of LOFAR has been driven by four fundamental astrophysical
applications: (i) The Epoch of Reionisation, (ii) Extragalactic
Surveys and their exploitation to study the formation and evolution of
clusters, galaxies and black holes, (iii) Transient Sources and their
association with high energy objects such as gamma ray bursts, and
(iv) Cosmic Ray showers and their exploitation to study the origin of
ultra-high energy cosmic rays.  In this conference the foreseen LOFAR
work on the epoch of reionisation has been covered by de Bruyn and on
cosmic ray showers by Falcke.

During this contribution we will first present the LOFAR project with
an emphasis on the challenges faced when carrying out sensitive
imaging at low radio frequencies.  Subsequently, we will discuss
LOFAR's capabilities to survey the low-frequency radio sky. Main aims
for the planned surveys are studies of $z>6$ radio galaxies, diffuse
emission associated with distant clusters and starbursting galaxies at
$z>2$.

}
   \maketitle
%
%________________________________________________________________
%
\section{Introduction}

During the last half century our knowledge of the Universe has been
revolutionized by the opening of observable windows outside the narrow
visible region of the spectrum. Radio waves, infrared and ultraviolet
radiation and X- and gamma rays have provided new and completely
unexpected information about the nature and history of the Universe
and have resulted in the discovery of a cosmic zoo of strange and
exotic objects. One of the few spectral windows that still remain to
be explored is at the low radio frequencies, the lowest energy extreme
of the accessible spectrum. LOFAR, the Low Frequency Radio Array, is a
large radio telescope that will open this ``Terra incognita''  to a broad
range of astrophysical studies.  In this contribution we will first 
briefly present the LOFAR telescope and the status of the LOFAR project. 
Subsequently, we will  give a short  overview of 
the scientific topics for which LOFAR is expected to make important contributions. 
Finally, the impact of LOFAR surveys as a tool to study the formation and 
evolution of clusters of galaxies, galaxies and AGN is discussed. 

\section{The LOFAR telescope} 

The design of the LOFAR telescope makes as much use as possible of modern computing technology. 
The main reason for this is that this is the only way to build a telescope  with  the needed effective aperture in  an affordable way. A very good example is the usage of phased array technology, which 
replaces the need for large mechanical dishes with a combination of simple receivers and a cluster of powerful 
PCs that together create the beams with which astronomers observe. A second good example is the correlator, which in the case 
of LOFAR is 
a state of the art supercomputer (an IBM blue Gene) instead of a custom made correlator. 

\begin{figure*}[t]
\centerline{
\includegraphics[width=8.8cm,angle=0]{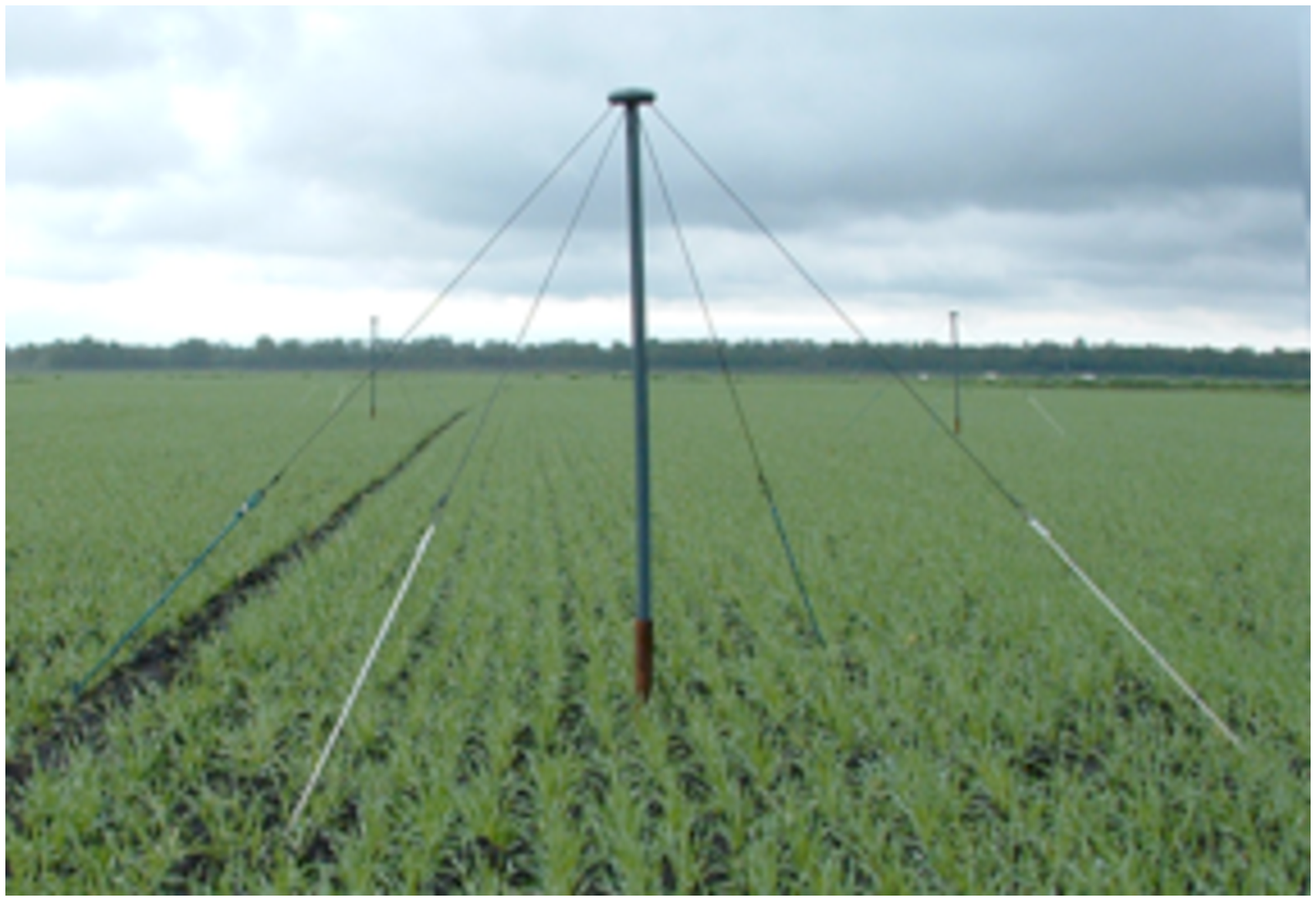}
\includegraphics[width=8.8cm,angle=0]{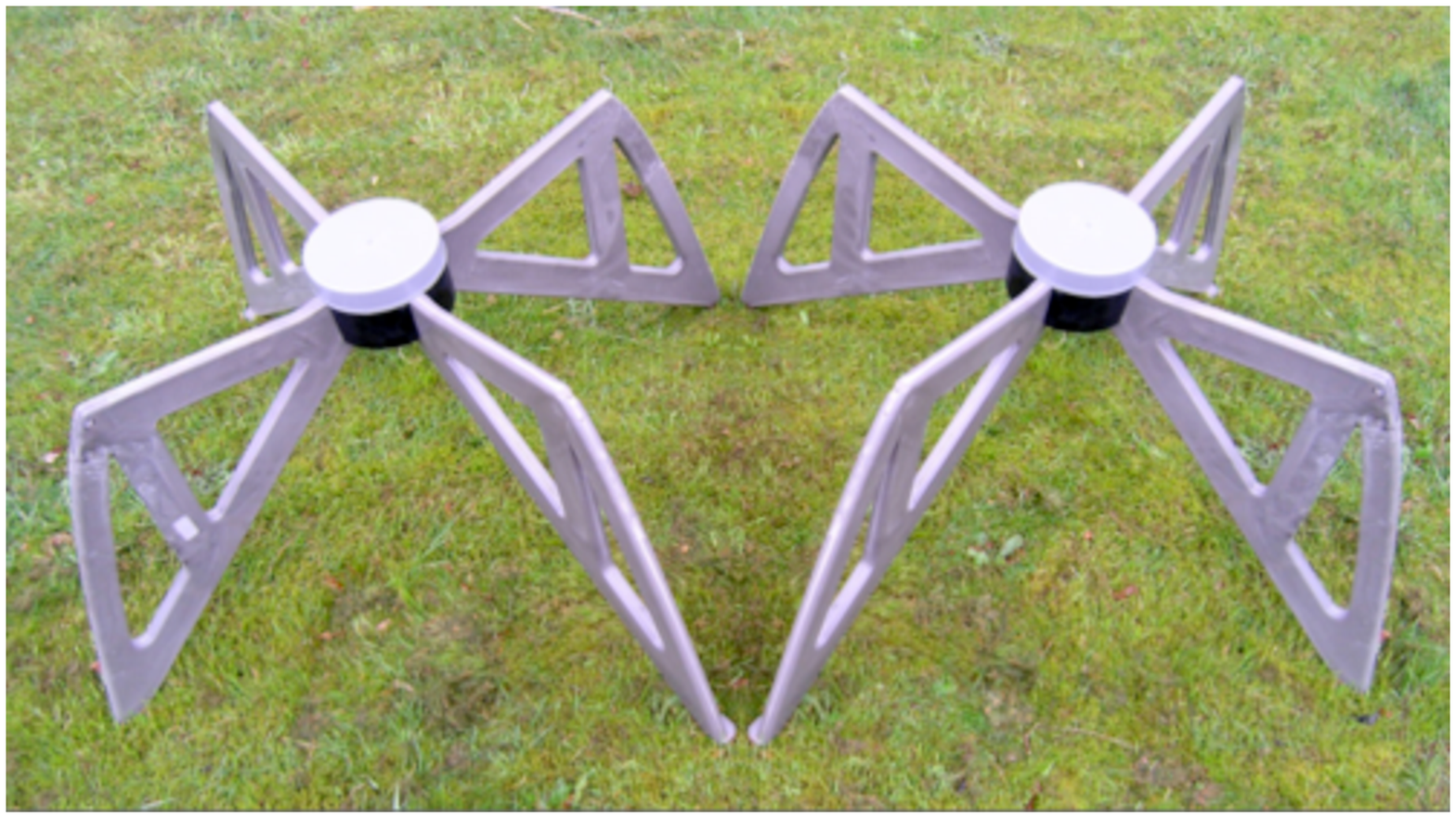}}
\caption{\label{antennas}
Pictures of  prototypes of the LOFAR antennas. ({\it left}) The low-freqency antenna, optimised 
for the 30-80  MHz frequency range. ({\it right})  The high-frequency antenna, optimised 
for the 115 - 240 MHz range. }
\end{figure*}

As just mentioned, the main receiver elements are relatively simple antennas. Two types of antennas are used to obtain  sufficient sensitivity over the nearly two decades that encompass LOFAR's observing bands.  The low frequency antennas are optimized for the $30 - 80$ MHz range and consist of dual-polarization dipole antennas (for pictures of proto-type antennas, see Fig. \ref{antennas}). Although at reduced sensitivity, the lowband antennas can observe below 30 MHz, down to the atmospheric  cutoff, which is -- depending on ionospheric weather conditions -- between 10 and 20 MHz. The high frequency antennas have their maximum sensitivity between 115 and 240 MHz. A picture of a proto-type is also  given in Figure \ref{antennas}.  Note that we are not planning to carry out sensitive observations between 90 and 115 MHz. The main reason for this is that the signals from a number of strong TV stations will prevent sensitive observations in this frequency range. 

About hundred low and hundred high frequency antennas will be placed in soccer-field sized fields to form one LOFAR station. The electric signals from the antennas will be transported to a series of electronic boards where appropriate phase delays will be applied so that station beams on the sky can be formed.   In the standard observing mode 8 beams are made simultaneously. 

\begin{figure*}[t]
\centerline{
\includegraphics[width=8.8cm,angle=0]{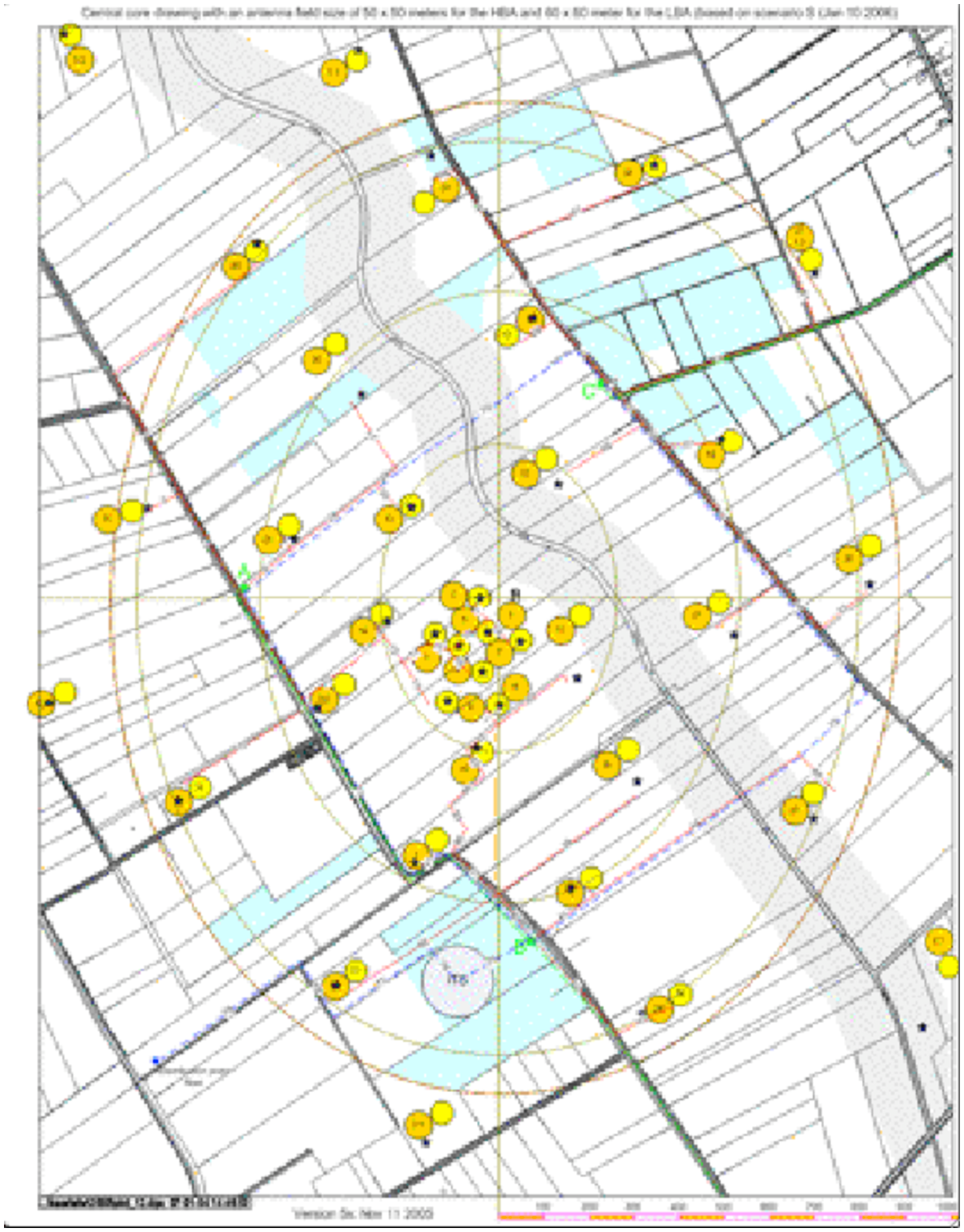}
\includegraphics[width=8.8cm,angle=0]{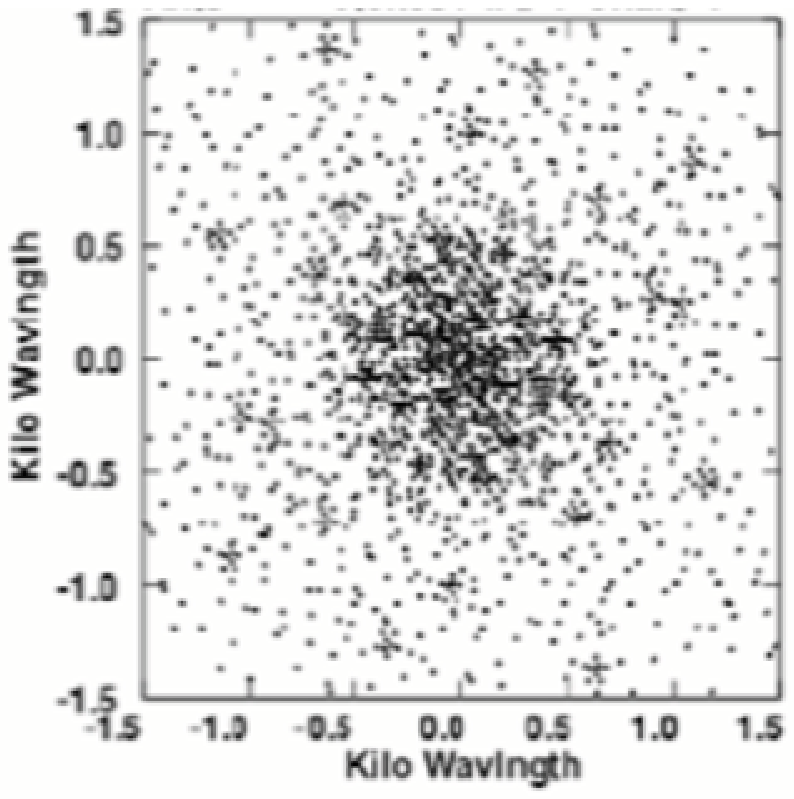}}
\caption{\label{layoutcore}
({\it left}) The location of the 32 LOFAR stations in the $2\times 3$ km$^2$ sized region near the municipality of Borgen-Odoorn in the Northern province of Drenthe. 
({\it right}) The instantaneous resulting UV coverage.} 
\end{figure*}

Over an area  with a diameter of 100 km,  77 stations will be build, with 32 stations spread over a  central $2\times 3$ km region. There are a number of reasons to have such a dense filling of stations in the central area. The two main scientific reasons are (i) to have sufficient sensitivity at angular scales of 
a few arc-minutes needed for the 21 cm lines studies of the epoch of reionisation and (ii) to allow 
for sensitive monitoring of the radio sky at relatively low angular resolution with the aim  of  detecting  and monitoring  radio transients on times-scales down to  1 second. The technical reason is related to calibrating the corrupting influence of the ionosphere on the visibility measurements. With such a sensitive central region, the calibration sources will give sufficient signal also on the longer baselines, so that phase  variations induced by the ionosphere can be corrected for on sub-minute time-scales. 
With land acquisition in the central area  almost complete, a final decision on the layout of the LOFAR core was taken in February 2006. Figure \ref{layoutcore} 
shows the layout of the central 32 stations.  Extensive simulations of both the instantaneous and longer integration UV-coverage have not just been instrumental in finding the optimum distribution of core stations, but have also guided the land acquisition process outside the core. 
The resulting excellent instantaneous UV coverage is given also in Figure \ref{layoutcore}.

\begin{figure*}[t!]
\centerline{
\includegraphics[width=6.0cm,angle=0,clip]{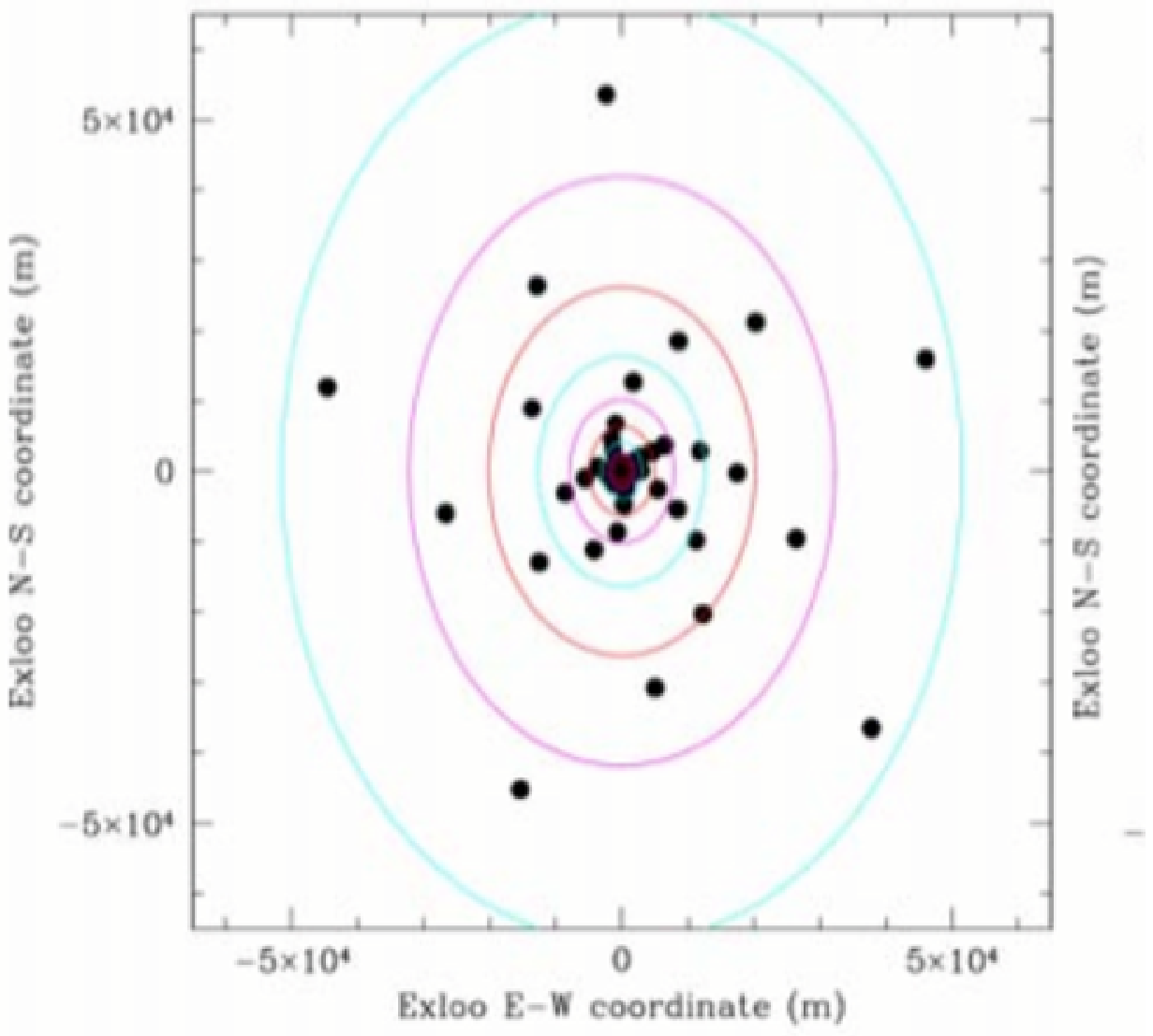}
\includegraphics[width=5.6cm,angle=0,clip]{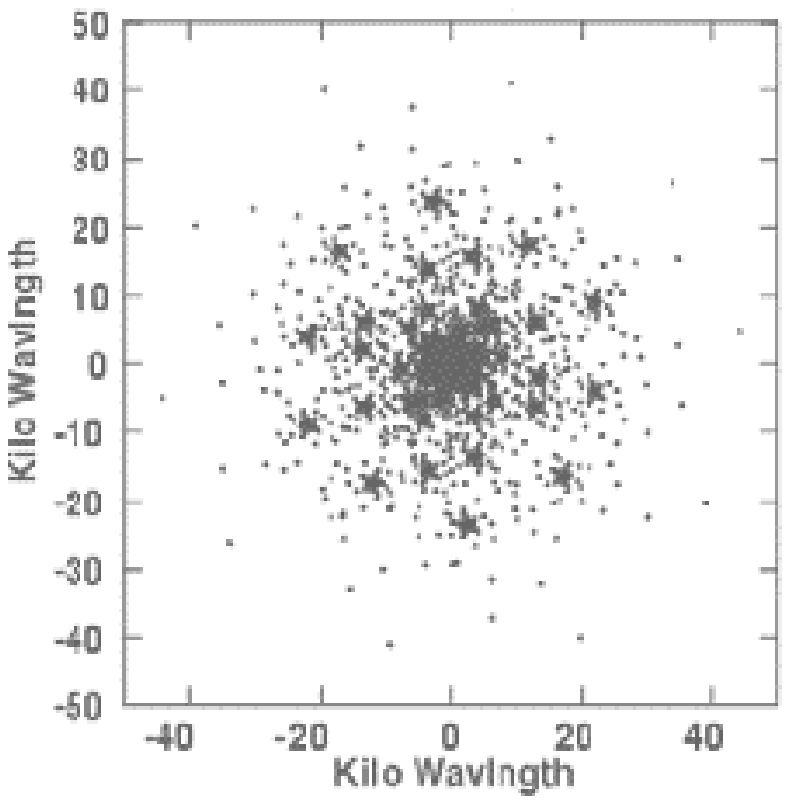}
\includegraphics[width=5.9cm,angle=0,clip]{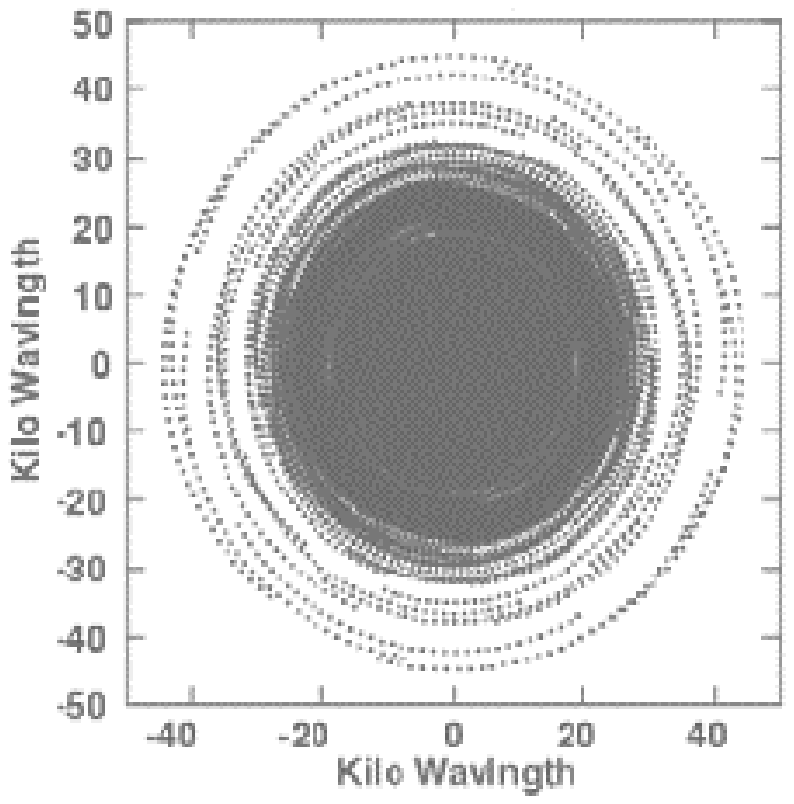}} 
\caption{\label{layoutlofar}
({\it left}) The layout of the 100 km LOFAR array. The resulting UV coverage is show 
for short integrations ({\it middle}) and long integrations ({\it right}). } 
\end{figure*}

The larger baselines of LOFAR will come from stations  placed in 9  consecutive rings, 
each containing 5 regularly spaced  stations. To ensure an excellent UV coverage for short and long 
integrations, the increment in distance from the center for each ring will follow a power-law
and the orientation of each subsequent ring will differ by roughly ten degrees. 
The resulting layout and  UV coverages are given in Figure \ref{layoutlofar}. 

Each of the stations will have at least a 3 Gb/s connection to the central processor. This central 
processor is an IBM Blue/Gene computer named `Stella'. With its $\sim 25$ Tflops capability 
combined with a high-throughput I/O system, it can correlate 32000 channels of 1 kHz each.
In the standard imaging mode, the resulting visibilities will be  average down to a spectral 
resolution of 10 kHz and a time resolution of 10 sec. 

An important milestone will be the delivery of the first Core Station (CS1) planned to be fully operational towards the end of 2006.  It will consist of the final prototype hardware needed to build a single LOFAR station. To facilitate experiments and tests, the 96 dual-polarization dipole antennas will be distributed over 4 core station fields. Half the antennas will be in a single field, the remaining 48 will be equally distributed over the other three stations. Although initial preparation for the roll-out of the array 
has started, only after successful testing of CS1 the full role-out will take place. The current schedule 
is that the last station for the hundred kilometer LOFAR should be in place at the end of 2008. 

It was  always the ultimate aim that LOFAR should have baselines up to at least four hundred kilometers. The main reason for this is that at the higher frequencies LOFAR will then have an angular  resolution of an arsec. This is clearly crucial for studies also involving data from optical, IR and X-ray 
telescopes, where the angular scales probed are often an arcsec or better.  Currently there is considerable interest in LOFAR among astronomers in Germany, the UK, France and Italy 
and consortia are being formed. Funding for  at least two stations in the UK
and three stations in Germany has almost been  secured. This is  clearly a very important step towards 
a fully functioning LOFAR array with the foreseen long baselines. 

\subsection{Challenges} 

Obtaining high quality maps with LOFAR faces a number of challenges and these need to be taken into account when designing
LOFAR survey observations.  Here we briefly  describe these challenges. 

\subsubsection{\label{rfi}Radio frequency interference (RFI)}

A major challenge for sensitive low frequency radio observations is radio frequency interference (RFI).
This is addressed in a number of ways. First, the design of LOFAR is such that due to the  usage of 12 bits A/D converters, the system should remain linear in the presence of strong RFI. Sources of interference can therefore be removed, with procedures similar to those used to remove the
impact of strong radio sources such as Cygnus A and Cassiopea A. Second, the system is capable of observing at high spectral resolution (1 KHz). Since RFI signals are often emitted in a narrow band, they will only affect a small fraction of frequency channels, presently estimated to be smaller than \hbox{10 \%.} Furthermore, RFI signals are often intermittent and due to the high time resolution with which LOFAR observes, they can be removed. 
Finally, the dominant noise source at these low frequencies is due to the high brightness temperatures of the sky. Faint RFI sources  from  for example  far away TV stations will
be a negligible contribution to the system noise  and will not add up coherently during longer integrations.

\subsubsection{Ionosphere}

For observations specially at lower LOFAR frequencies and longer baselines, the ionosphere 
severely corrupts the quality of the visibility measurements, making 
calibration of the UV data particularly challenging. On the basis of  a simple 
parametrization of  the ionospheric induced phase fluctuations, 
phase corrections will  be derived  from the data as a function of location of the antennas in the array  and on viewing direction. Furthermore, these corrections need to be updated every few seconds. Currently the LOFAR project is working on implementing such a calibration scheme.
The scheme has been named `Peeling' (Noordam 2004). \nocite{noo04} It  iteratively finds the brightest sources  within 
a beam and subsequently determines the ionospheric corrections towards 
these sources. These corrections are then jointly used to calibrate the data. 
Although it is  expected that under  normal observing conditions, 
well calibrated maps can be produced, residual variations of the shape of the point-spread-function over the map 
will remain. These need to be taken into account during the process of extracting and characterizing 
sources.

\subsubsection{Computational resources}

Because of the inherent large-sky nature of each LOFAR observation, the computational and data transport resources that will be needed to produce standard LOFAR maps
are formidable. Current estimates indicate that to produce well calibrated maps from the raw visibility measurements, a cluster with at least a 1000 2\,GHz processors are needed. As presently planned, standard LOFAR maps will be made at the LOFAR Science Centre using a Linux cluster of 1000 parallel processors from visibility data outputted by the Blue-Gene/Stella correlator. 

\subsection{Performance} 
The resulting performance for the 100 km LOFAR array is given in Table \ref{lofar}
(see also http://www.lofar.org/).
The resolution is  defined as $1.22\times\lambda/D$, where $\lambda$ is the observing wavelength and D is the array size.  The sensitivity is based on an estimate of the effective aperture of the LOFAR stations
and a simple extrapolation of the sky brightness. Note that LOFAR has been designed 
such that the noise due to the high sky brightness temperatures is the dominant noise source. 
The size of a LOFAR station, taken as 50 meters,
determines the angular size of the 
field of view of a station, i.e. the LOFAR station beam. 
The size of the station  beam together with the number of beams that 
are formed determines the 
fraction of the sky that can be observed instantaneously. For the standard observing mode using the 
whole array  8 beams are formed. For the central cores, 32 beams can be formed 
allowing for covering efficiently a large fraction of the sky. 
The number of station beams needed to cover the entire sky accessible to LOFAR above declination $\delta = 0^{\circ}$
is also given in Table \ref{lofar}.

\begin{table}
\caption{\label{lofar}
Main scientific capabilities of the LOFAR array, see also www.lofar.org. }
\begin{center}
\begin{tabular}{cccccc}
\hline \hline 
$\nu$&  $\lambda$  &Sensitivity &  Resol. &   Primary & Number \\
  &&&&beam& beams \\
    MHz        & m & mJy/beam  &    arcsec   &  degree &         \\
(1)&(2)&(3)&(4)&(5)&(6)\\
\hline
    15&  20.0&11&    50.3&   22.9&     50\\
    30&  10.0& 2&    25.2&   11.5&    200\\
    60&   5.0& 1.65&    12.6&    5.7&    800\\
    75&   4.0& 1.30&    10.1&    4.6&   1250\\
   120&   2.5& 0.070&     6.3&    2.9&   3200\\
   150&   2.0& 0.065&     5.0&    2.3&   5000\\
   200&   1.5& 0.063&     3.8&    1.7&   8888\\
   240&   1.2& 0.076&     3.1&    1.4&  12800\\
\hline
\end{tabular}
\end{center}
Notes for columns: (3)
Point source sensitivity given as the rms map noise for 1 hour integration time, 2 polarizations and 4 MHz bandwidth, (5) (Primary) beam size  calculated for a 50 meter station, (6) Number of independent beams needed to cover  $2\pi$ sterradian.
\end{table}

\section{Science} 

LOFAR will impact a broad range of astrophysics, ranging from cosmology to solar system studies. The  4 key areas that have driven the design of LOFAR are: 

\subsection{The Epoch of Reionisation} 
One of the most exciting applications of LOFAR will be the search for redshifted 21cm line emission from the Epoch of Reionisation (EoR). LOFAR will address a number of key questions related to the EoR, including: 

\begin{enumerate} 
\item What is the redshift range in which the bulk of the neutral hydrogen became ionized? Can a single ÔRedshift of ReionisationÕ be identified or defined or are there multiple phases of reionisation?
\item What are the characteristics of the spatial distribution of heated and still cold IGM and how do these evolve during the era of reionisation? 
\item Which objects (Pop III stars, galaxies, quasars) or processes are responsible for re-ionizing the Universe?\end{enumerate} 

During this conference, an overview of the LOFAR 
EoR project has been given by de Bruyn and we refer to his contribution
for a detailed account. 

\subsection {Extragalactic Surveys and their exploitation to study the formation and evolution of clusters, galaxies and black holes} 

One of the most important applications of LOFAR will be to carry out large-sky surveys. Such surveys are well suited to the characteristics of LOFAR and have been designated as one of the key projects that have driven LOFAR since its inception.  Such deep LOFAR surveys of the accessible sky at several frequencies will provide unique catalogues of radio sources for investigating several fundamental questions in astrophysics, including the formation of massive black holes, galaxies and clusters of galaxies. Because the LOFAR surveys will probe unexplored parameter space, it is likely that they will discover new phenomena. 
In Section \ref{survey} we will further discuss this project. 

\subsection{
Transient Sources and their association with high energy objects such as gamma ray bursts} 

LOFAR's large instantaneous beam will make it uniquely suited to efficiently monitor a large fraction of the sky, allowing a sensitive unbiased survey of radio transients for the first time. Averaging of the data will provide information on a variety of time scales, ranging from seconds to many days. The resolution attained will be sufficient for the crucial task of rapid optical and X-ray identifications. Table
\ref{trans} gives an brief overview of the classes of object known or expected to exhibit variable radio emission. This overview has been 
made by the team that is leading the efforts for the transients project. This team consists 
of Fenders (Southampton), Wijers (Amsterdam), Braun (ASTRON) and Stappers (ASTRON/Amsterdam).
Also indicated are the variability time-scales, the number of objects/events that are expected to be observed per year and an estimate of the distances to which these objects can be seen. 

\begin{table*}
\caption{\label{trans} Predicted detection numbers of transient sources}
\begin{tabular}{llcc}
\hline
\hline 
Class of object 	&Time-scale	& Expected / year&	Maximum Distance \\
\hline
GRB afterglows  &	months	& $\sim 100$	&Observable universe\\
LIGO Events	&msec / hours	&a few ?&	Observable universe\\
Radio Supernovae &	days / months	 &$\sim 3$	&100 Mpc\\
Intermediate mass &BH	days 	&$1-5$	&30 Mpc\\
Flare Stars 	&msec / hours	 &$100-1000$&	1 kpc\\
Exo-planets 	&min / hours	& $10-100$	&30 pc\\
\hline 
\end{tabular}
\end{table*}%

\subsection{Cosmic Ray showers and their exploitation to study the origin of ultra-high energy cosmic rays}

LOFAR offers a unique possibility in particle astrophysics for studying the origin of high-energy cosmic rays (HECRs) at energies between $10^{15} - 10^{20.5}$ eV. Both the source origin and the acceleration processes of these particles are unknown. Possible candidate sources of HECRs are shocks in radio lobes of powerful radio galaxies, intergalactic shocks created during the epoch of galaxy formation, so-called Hyper-novae, Gamma-ray bursts, or decay products of super-massive particles from topological defects, left over from phase transitions in the early universe.

The uniqueness of LOFAR as a cosmic ray (CR) detector was discussed
quantitatively in for example Falcke and Gorham 2003 \nocite{fal03} 
and lies in its capacity to measure:
\begin{itemize} 
\item the composition of CRs via the Gerasimova-Zatsepin effect due its
remote stations,

\item point sources of high energy CRs inside the galaxy via the detection
of clustering due to energetic ($>10^{18}$ eV) neutrons,

\item neutrinos that have collided with the lunar surface regolith,

\item very accurate positions on the sky of better than $<<1^\circ$, surpassing
any other CR detection method.
\end{itemize} 

For a detailed account of the prospects that LOFAR offers for HECR studies we refer to the 
contribution of Falcke to this conference. 

\subsection{and even more science}

These 4 science projects  are only a subset of the 
astronomical topics that LOFAR will adress. Important studies that 
can and will be conducted include high time and angular resolution  observations of the 
sun and Jupiter, and detailed studies of Galactic sources such as 
supernova remnants, HII regions, and pulsars. Also important data will be obtained 
for studying the dynamic behavior of the ionosphere on a variety of time and angular scales. 

Although outside the scope of this presentation, it is interesting to note that the 
LOFAR network and computing facilities will also  be used for non-astronomical investigations. 
For these investigations, new sensors are coupled to the LOFAR system. Currently,
sensors that have already been developed are for applications related to agriculture and geophysics.

\section{Extragalactic Surveys and their exploitation to study the formation and evolution of clusters, galaxies and black holes} 
\label{survey} 

During this conference de Bruyn discussed the LOFAR project on the EOR and 
Falcke reviewed the status of  LOFAR's CR project. Here we will elaborate  on the 
LOFAR project to carry out large surveys of the low-frequency sky.  
The survey project as currently planned will survey the entire accessible
sky at frequencies of 15, 30, 60 and 120 MHz and a few hundred
square degrees at 200 MHz. The combination of these surveys has been
selected to optimize LOFARÕs impact in three important areas of
extragalactic astrophysics:

\subsection{The formation of massive galaxies, clusters and black holes using the most distant radio galaxies as probes} 

With radio luminosities of up to 5 orders
of magnitude greater than our own galaxy,
luminous distant radio galaxies are
unique cosmic probes. 
First, they are amongst the
most energetic objects in the Universe, with radio powers of up to
10$^{28}$ W Hz$^{-1}$, and are therefore important laboratories for probing
high-energy phenomena in the early Universe. It has
been suggested that the black holes may be formed before the galaxies
themselves and that the jets produced by the putative black holes
stimulate star formation and play an important role in forming the
first galaxies (Silk and Rees 1998). \nocite{sil98}
Secondly, they are among the brightest known galaxies in the early
Universe (De Breuck et al. 2002)
\nocite{bre02a} and the likely
progenitors of dominant cluster galaxies (Pentericci et al. 1997;
Pentericci et al. 1999; Miley et al. 2004). \nocite{pen97a,pen99a,mil04}
Thirdly, powerful radio galaxies  are often surrounded by significant galaxy overdensities,
whose structures have sizes of a few Mpc and velocity dispersions of a
few hundred km s$^{-1}$ (e.g. Venemans et al. 2002 and references therein). 
\nocite{ven02} The inferred
masses of these structures  are several
$\times 10^{14}$ M$_\odot$, consistent with them being the ancestors
of rich clusters (protoclusters). Such radio selected protoclusters
have been established out to $z \sim 4.1$ (Venemans et al. 2002, Miley
et al. 2004; Overzier et al. 2006), and a probable case has been found
at $z \sim 5.2$ (Venemans et al. 2004b; Overzier et al. 2006).
\nocite{ven04,ove06}

The most efficient method for finding extremely distant radio galaxies uses an empirical correlation between radio spectral steepness and distance (R\"ottgering et al. 1997, de
Breuck et al. 2000).
\nocite{rot97a,bre00c}  Using this method, LOFAR will efficiently pick out radio galaxies at larger distances than possible with present radio telescopes. The possibility of detecting such objects close to or before the epoch of reionisation is particularly exciting. Not only would the discovery of such objects provide important constraints on how and when massive black holes are formed, but $z > 6$ radio galaxies would also allow detailed studies of the ISM at these high redshifts through redshifted 21 cm absorption studies. 

LOFAR surveys could detect and determine low frequency radio spectra for more than 50 million radio sources. Simple extrapolations of spectra and redshift statistics from existing radio and optical surveys (e.g. Jarvis et al. 2001; Fan et al. 2004)
\nocite{jar01b,fan04}
suggest that such LOFAR surveys will contain several hundred radio galaxies with redshift $z > 6$, i.e. located at distances larger than the most distant radio galaxy presently known. 

The objects with extreme spectra from the LOFAR surveys will be unique targets for follow-up with large optical and infrared telescopes. After their redshifts have been  measured, 
multi-wavelength studies of such LOFAR distant radio galaxies will provide information about the formation of massive galaxies and links between nuclear activity and star formation. Since distant radio galaxies have been shown to pinpoint proto-clusters, studying the environments of these objects constrain the formation of clusters at the earliest epochs.

\subsection{Intercluster magnetic fields using diffuse radio emission in galaxy clusters as probes}

Clusters often contain diffuse radio sources that provide unique
tracers of the intercluster magnetic fields (e.g.  Govoni \& Feretti
2004).  \nocite{gov04} Comparison of radio and X-ray data show that
the radio sources are shaped by the dynamics of the gas in which they
are embedded. Their large extent, low surface brightness, and steep
spectra make these cluster radio sources difficult to study with
conventional facilities, such as Westerbork and the VLA. Because of
their steep spectra, LOFAR will be able to detect and study such radio
sources in the many tens of thousands of clusters up to redshifts of
$\sim 2$. These should also be detectable using the XMM-Newton X-ray
telescope, the Planck satellite, and the Sloan Digital Sky
Survey. Studies of LOFAR diffuse cluster sources will be used to (i)
probe the effects on the dynamics of cluster gas due to shock waves
produced by cluster mergers, (ii) determine the origin of cluster
magnetic fields, and (iii) measure the occurrence and characteristics
of diffuse cluster radio sources as a function of redshift with the
aim of constraining physical models for the origin of these sources
and the evolution of intercluster magnetic fields.

Ensslin and R\"ottgering (2002; ER02 hereafter) \nocite{ens02b} 
give a simple prediction of the local and higher
redshift radio halo luminosity function  on the basis of (i) an
observed and a theoretical X-ray cluster luminosity function,
(ii) the observed radio--X-ray luminosity correlation  of galaxy
clusters with radio halos, and  (iii) an assumed fraction of $
\frac{1}{3}$ galaxy clusters to have radio halos as supported by
observations.

For the evolution of clusters,  $z \sim 1$
corresponds to a crucial epoch, -- the transition epoch between clusters with properties
similar to local clusters
and unrelaxed proto clusters within which the hot central cluster gas is still being accumulated.
An  important goal for LOFAR therefore will be to obtain a substantial sample of radio halos at a redshift  of about  unity.
On the basis of the ER02  models, and optimizing depth versus area
for the  various frequency bands, it seems that a deep all sky survey at 120 MHz
is optimum for this purpose. 
Such a  survey might 
yield 10,000 radio halos up to redshifts of unity, with about 100 halos in the redshift range $0.8 < z <1.2$.

\subsection{Star formation processes in the early Universe using starburst galaxies as probes} 

Because of the large fields of view, deep surveys at the higher LOFAR frequencies will detect unprecedented numbers of star-forming galaxies with star formation rates of tens of solar masses per year at an epoch at which the bulk of galaxy formation is believed to occur. Since the ratio of radio flux to sub-mm flux is a sensitive redshift indicator, LOFAR surveys, in combination with data from new far-IR and (sub-)millimeter facilities such as the SCUBA-2 sub-mm imaging array, the Spitzer and Herschel satellites and the ALMA (sub-)mm array, will  provide distances and thus facilitate a census of the cosmic star-formation history, unhindered by the effects of dust obscuration.

Carilli and Yun (1999) \nocite{car99} give a relation between the
starformation rate and 1.4 GHz flux density for  starburst galaxies
as a function of redshift.  Following Condon (1992) \nocite{con92} we  used
a spectral index $\alpha =-0.8$  to  extrapolate this relation to lower
frequencies and/or high redshifts.  It is not clear whether this choice
is entirely appropriate.  Especially at the lower frequencies free-free absorption due to HII regions will flatten the spectra, while at
high redshifts Compton losses off the microwave background will steepen
the spectra.

An important aim is to probe in the LOFAR bands galaxies with  representative SFRs for the
$z > 2 $ galaxy population, taken to be 10 M$_\odot$/yr. 
Most  favorable seems  the 200 MHz
band, where deep observation allow such objects to be detected up  to $z\sim 2$. 
Such deep observations will also allow for the detection of 
extreme starbursts at much higher redshift than
presently is possible. While the current WSRT/VLA  detection limit for
100 M$_\odot$/yr starbursts is $z\sim 2$, the proposed very deep LOFAR 
surveys can probe back in time to redshifts of
$z\sim 7$ for such starbursts. 
 
The resulting samples of star forming galaxies will be used to study
star formation properties as a function of environment  and galaxy type (LB galaxies,
Ly$\alpha$ galaxies, red galaxies, submm galaxies). To be able to carry out such studies, sufficient
area needs to be covered.
For environmental studies, it is important that a range of environments up to redshifts of 3 are probed.
Ensembles of galaxies at $1<z<3$ that can be regarded as the progenitors of COMA cluster are rare. Simple  estimates give a few per 5 square degree. Hence, to obtain a sample of 100 proto-clusters,
250 sq  degrees need to be observed.

\begin{figure*}[t!]
\centerline{
\includegraphics[width=16cm,angle=0]{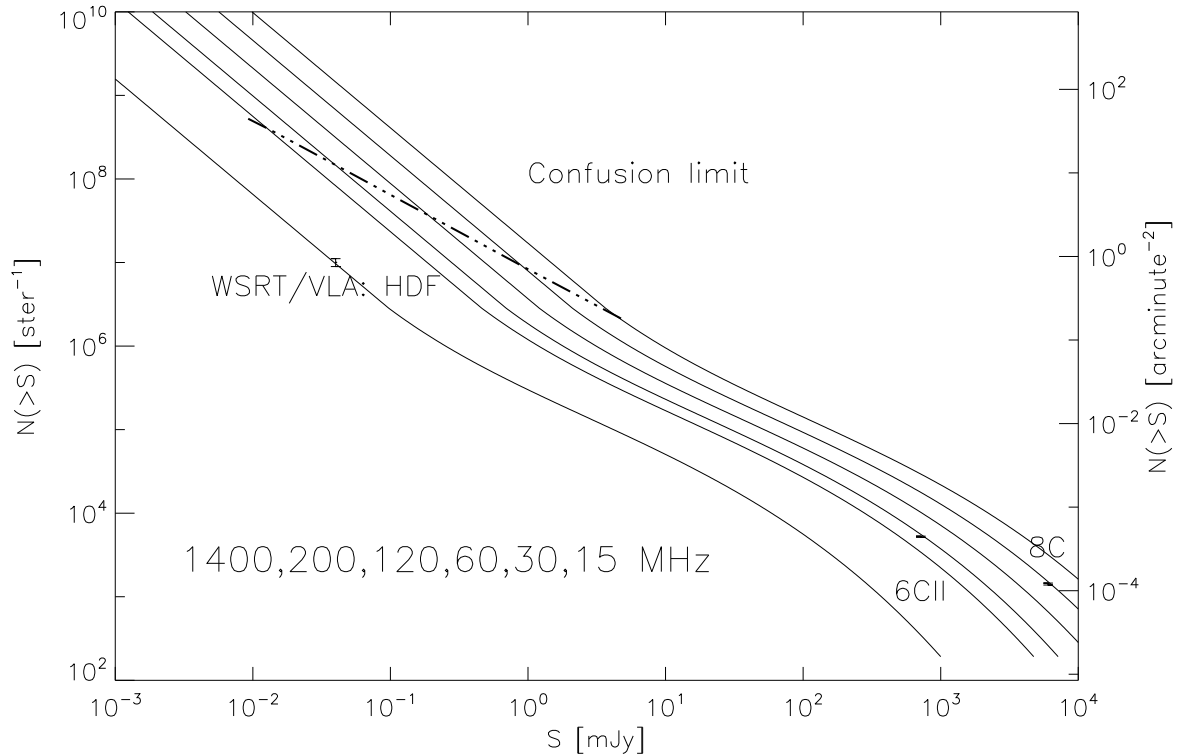}}
\caption{
\label{sourcecounts} Extrapolated radio source counts on the basis of 1400 MHz source counts  from
Katgert et al. (1988) and Richards (2000). The dot-dashed line indicates the
confusion limit for a 100 km LOFAR.  This limit is  defined as the source density at which there is on average 0.1 source per synthesized beam.
}
\end{figure*}

\subsection{A 15 MHz survey: Exploring unknown parameter space}

One of the most exciting aspects of LOFAR is its ability to explore unchartered
parameter space. There are 2 new areas of LOFAR parameter space of interest for serendipitous
discovery. First, the unprecedented size of its instantaneous field of view make it a unique
synoptic telescope for detecting new classes of transient sources. Short-term transient sources
will be searched for under the auspices of the transient project. 
The second unchartered parameter space that is most important for planning the surveys is at frequencies
$< 30$ MHz. These very low frequencies are the extreme low-energy end of the observable electromagnetic spectrum
and are virgin territory for observations. This is the LOFAR frequency regime with the highest probability of serendipitous discovery. It probes radiation mechanisms that are not observable at higher radio frequencies, such as coherent plasma emission. The bright very low-frequency sky is significantly different to the radio sky normally studied. For example, the brightest objects are the Sun and Jupiter with their very steep spectra ($\nu^{-3.5}$) between 20 and 40 MHz.
Deep searches for highly polarized low-frequency sources will be another new area of parameter space that LOFAR can penetrate. We note  that before pulsars were known, the crab pulsar was observed as a highly polarized ultra-steep spectrum continuum source at 38 MHz.  

In principle, the amount of time needed to carry out an all sky survey at
frequencies below 30 MHz is limited. In practice, it might be that for only
a very small fraction (5 \% ?) of the time the ionosphere is stable enough to allow good data
to be obtained. We plan to carry out
such ultra low frequency  surveys at 15 MHz.  This  frequency is a compromise
between not being  too close to the ionospheric cutoff below which the atmosphere is
opaque and being significantly different from the 30  MHz survey.

\subsection{Additional projects with LOFAR surveys}

Although the main drivers of LOFAR surveys are the above-mentioned topics, LOFAR surveys have unique characteristics that will impact a broad range of astrophysics.
Additional topics for which LOFAR will make important contributions include:

\begin{itemize}

\item
Large scale structure of the Universe.

Besides using LOFAR radio galaxies to pinpoint distant protoclusters, large-scale structure can be studied in a number of ways. These include: (i) radio imaging of galaxy filaments connecting clusters of galaxies as an important probe of shocks in a magnetized warm/hot intergalactic medium, (ii) searches for strongly lensed radio sources making use of LOFARÕs capability of mapping large fields at high angular resolution, (iii) due to the large number of expected sources, the cosmological clustering of radio sources as a function of redshift, type of host and environment can be studied in unprecedented detail and (iv) Thomson scattering halos expecting to occur around some clusters of galaxies.

\item
Physics of the origin, evolution and end-stages of radio sources.

Studies of giant radio sources constrain aspects of radio source physics, 
such as the lifetimes of the AGN process and the properties of the medium 
in which such sources are embedded.  Due to LOFARÕs unprecedented surface 
brightness sensitivity at low radio frequency, large numbers of giant radio 
sources should be detected with sizes well in excess of 1 Mpc. These, as
well as relic AGN emission,  will be a powerful tool for constraining the 
late stages of radio source evolution

The long wavelength range of LOFAR will provide unique information on absorption processes in compact extragalactic radio sources and thereby constrain the nature of the absorbing medium. An important challenge is to accurately measure the low frequency spectrum with the aim of distinguishing between synchrotron self absorption and free-free emission.

\item
Magnetic field and Interstellar medium in nearby galaxies.

 For nearby galaxies, LOFAR will enable spectral mapping at low frequencies, thereby providing unique information about absorption and delineating the spatial distribution of the warm ISM and galactic magnetic fields.

\item
Galactic sources.

 Due to the large fields of LOFAR, the proposed surveys will contain important information on many classes of galactic sources that are each worthy of specific study. These include (i) supernova remnants, (ii) HII regions, (iii) exo-planets and (iv) pulsars.

\end{itemize}

\subsection{Source counts}

\begin{table*}[t!]
\caption{\label{prop} Proposed LOFAR surveys }
\begin{tabular}{ccccccll}
\hline
\hline
$\nu$& Flux$^\dagger$  & Area & Source&Number &Int. time$^\ddagger$ &  Total&Main aim  \\
           &         density   &&density&Sources              &  &4 beam \\
MHz   &          mJy         &&  arcmin$^{-2}$   &&hour&  years \\
(1)      &   (2)              &  &         (3)&(4)&(5)&(6)\\
\hline
    15& 4.7&   2$\pi$  sr &    0.2&     $1.3 \times 10^7$&      48&    0.07 &New parameter space   \\
    30& 1.0&   2$\pi$ sr  &    0.7&     $5.4\times 10^7$&      38&    0.22 &$z > 6$ radio galaxies  \\
    60& 1.0&  2$\pi$  sr &    0.3 &   $2.2\times 10^7$ &    24 &    0.56 &Spectral information   \\
   120& 0.043&  2$\pi$ sr   &   11.6&     $8.6\times 10^8$&      23&    2.17 &Distant cluster halos\\
&&&&&&                                      &$z > 6$ radio galaxies   \\
   200& 0.014&  250 deg$^2$ &   32.2&    $ 3.0\times 10^7$&   191&    0.61 &Distant starbursts   \\
\hline
\end{tabular}
\space
\begin{quote}
Notes:\\ 
$^\dagger$ The specified flux limits given are 3-sigma values. \\
$^\ddagger$ The integration time needed using one LOFAR beam of 4 MHz 
to reach the 
specified flux limit.
The surveys will be carried out in several passes to detect sources that vary on timescales of weeks to years.\\
\end{quote}
\end{table*}

To plan an optimum set of LOFAR surveys, it is important to estimate the number of expected
sources and their flux distribution at the LOFAR frequencies.
Sensitive observations at low frequencies do not exist - which is  one of the
reasons to build LOFAR. However, we can make an estimate of the number of sources that LOFAR will observe
by extrapolating from data from surveys at higher frequencies. 
As a starting point we  have taken the Euclidean normalized source
counts at 1.4 GHz for the flux  density range 0.1 mJy up to 10 Jy
as parameterized by Katgert et al. 1988. \nocite{kat88} These counts are 
extrapolated to faint limits adopting
the slope for the source counts below 0.1 mJy as measured by 
Richard (2000) \nocite{ric00} using a very deep 1.4 GHz survey
centered at the Hubble Deep Field. 
Using a spectral index of $\alpha = -0.8$, the expected LOFAR source counts
were calculated at the relevant frequencies and plotted in Figure \ref{sourcecounts}.

A survey is defined to be confusion limited if faint ``undetected'' sources with flux densities near the
map noise level contribute significantly to the overall noise in the map. In this case longer integration times
will not lead to a substantial increase  of signal to noise of sources in the radio maps.
Especially for  a 100 km LOFAR, a very sensitive instrument with modest angular resolution,
this is an issue.
We will define the confusion limit as the surface density at which
there is on average 0.1 source per synthesized beam.
The values for this limit as a function of frequency are plotted
as the dashed-dotted line in Figure \ref{sourcecounts}.
Whereas for a 100 km LOFAR the deeper continuum surveys will tend to be confusion 
limited, this is much less of an issue for a $>400$ km LOFAR.

\section{Set of Proposed surveys}

The discussion of the various requirements for the main scientific drivers leads to 
a coherent set of surveys at 15, 30, 120 and 200 MHz. 
However, this leaves a large gap in spectral coverage between
the 30 and 120 MHz surveys. This gap is very problematic for a proper
mapping of the low frequency spectral properties of many types of radio sources.
Especially damaging would this be for studies of very extended objects
such as  giant radio sources, nearby galaxies and radio halos and relics, where  large spatial variation
of spectral properties are expected and can be used as a powerful tool to constrain physical mechanisms within these sources.
Furthermore, samples of low frequency peaked spectra can only be  well selected
if their spectra are sufficiently  sampled.

We therefore plan  to augment the suite of surveys with one centered at 60 MHz.
60 MHz was selected since it is the harmonic mean of the 30 and 120 MHz surveys, giving
optimal spectral coverage. The depth is dictated by our aim 
of detecting all the 30 MHz sources in the planned  survey with a spectral index $\alpha > 0$ between 30 and 60 MHz.

The surveys as 
we presently foresee them are   summarized in Table \ref{prop}. 
We note that the  information in this table is to illustrate our current thinking of what would 
an optimum use of  survey time for  a 100 km LOFAR. The final survey  product might well be different 
since it will take into account (i) the measured in stead of predicted capabilities of the instruments, (ii) baselines larger than 100 km,  
and (iii) further scientific input from a broad community.

\begin{figure*}[t!]
\centerline{
\includegraphics[width=12cm,angle=+90]{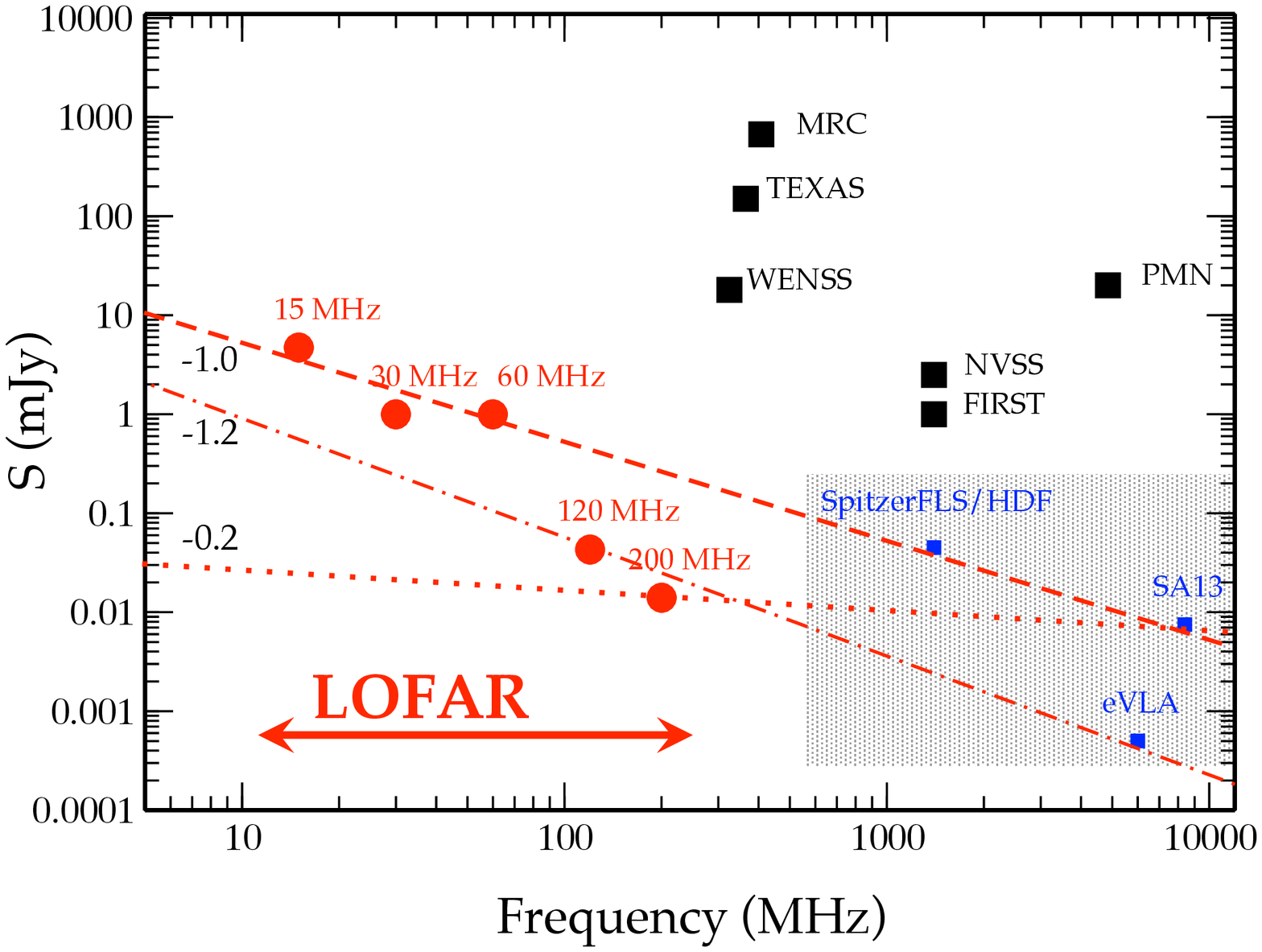}}
\caption{\label{other}
Flux limits of the proposed LOFAR surveys (as described in Table \ref{prop}) compared
to other planned and existing radio surveys.  The black squares represent
existing large (but shallow) surveys.  The grey area represent the typical
flux limit reached by the deep but small-area (at most few squares degrees)
surveys now available.  Few examples are illustrated (blue squares):
HDF (VLA Richards et al. 2000; WSRT Garrett et al. 2000), the Spitzer First Look survey
(WSRT, Morganti et al.  2004), the SA13 field (VLA, Fomalont et al. 2002).  
200 hour eVLA continuum surveys at 1.4 and 6 GHz  are also added for
reference.
The lines represent different power-laws ($S\sim \nu^{\alpha}$, with $\alpha=
-1.0, -1.2, -0.2$) to illustrate how, depending on the spectral indices of the
sources, the LOFAR surveys will compare to others surveys.  The deeper
LOFAR surveys (at 120 and 200 MHz) will be able to see the relatively flat
spectrum sources detected even in the deepest available fields.  For sources
with spectral index $-1.0$, also the 15, 30 and 60 MHz surveys will have
comparable sensitivity to the deepest field done so far.  For sources with
spectral index steeper than $-1.2$, the 120 and 200 MHz deep LOFAR surveys will
have a comparable sensitivity to the ultra deep eVLA survey.  It is, however,
important to remember that the LOFAR surveys will cover a $2\pi$ area of the sky
(or at least 250 deg$^2$ at 200MHz). } 
\end{figure*}

The proposed surveys will not only be unique due to their low frequencies,
but are also very important due to their depth as compared to other surveys.
To illustrate this we have plotted the limits of a number of surveys in Figure \ref{other}
as a function of frequency together with the proposed surveys.
It is clear that LOFAR performs 2-3 orders of magnitude better than
any other existing large  sky radio survey. Even for flat spectrum sources the depths of the higher frequency
LOFAR surveys are  comparable to for example the 8GHz SA 13 survey
(Fomalont et al. 2002).

\nocite{ric00,gar00,mor04,fom02}

\section{Conclusions}

The 100 km LOFAR will be a very powerful instrument for a broad range of studies. It will have a significant impact on at least the 4 major areas of astrophysics 
that are discussed in this proposal. What in the end might turn out to be even more important 
is that this new machine really probes new parameter space, which will
hopefully lead to many unexpected  discoveries. Especially for extragalactic 
studies the extension of baselines well beyond 100 km is very important. This would 
clearly dramatically increase LOFAR's impact on studies of distant radio galaxies, 
clusters sources and starforming galaxies.

\begin{acknowledgements}
The authors are very grateful  to the ASTRON and LOFAR staff, who have been 
and are still crucial to make of LOFAR a well functioning low-frequency observatory. 
HR would like to thank Chas Beichman and George Helou for their warm welcome 
to the Michelson and Spitzer science centers where  this  contribution was written. 
\end{acknowledgements}

%\bibliography{/Users/Huub/texinputs/huub}
%\bibliographystyle{/Users/Huub/texinputs/aa}

\end{document}